\documentclass[aps,prb,twocolumn,groupedaddress,showpacs]{revtex4}
\usepackage{epsfig}
\usepackage{graphicx}

\def  \bsig    {\mbox{\boldmath$\sigma $}}
\def  \btau    {\mbox{\boldmath$\tau $}}

\begin{document}

\title{Exchange interaction of magnetic impurities in graphene}

\author{V. K. Dugaev$^{1,2,3}$, V. I. Litvinov$^4$, J. Barna\'s$^{5,6}$}

\affiliation{$^1$Department of Mathematics and Applied Physics, University
of Technology, Al. Powsta\'nc\'ow Warszawy 6, 35-959 Rzesz\'ow,
Poland}
\affiliation{$^{2}$Departamento de F\'isica and CFIF, Instituto Superior
T\'ecnico, Av. Rovisco Pais, 1049-001 Lisbon, Portugal}
\affiliation{$^3$Frantsevich Institute for Problems of Materials Science,
National Academy of Sciences of Ukraine, Vilde 5, 58001 Chernovtsy, Ukraine}
\affiliation{$^4$WaveBand/Sierra Nevada Corporation, 15245 Alton
Parkway, Suite 100, Irvine, CA 92618, USA}
\affiliation{$^5$Department of Physics,
Adam Mickiewicz University, Umultowska 85, 61-614 Pozna\'n,
Poland}
\affiliation{$^6$Institute of Molecular Physics, Polish
Academy of Sciences, Smoluchowskiego 17, 60-179 Pozna\'n, Poland}

\date{\today }

\begin{abstract}

The indirect exchange interaction between magnetic impurities
localized in a graphene plane is considered theoretically, with
the influence of intrinsic spin-orbit interaction taken into
account. Such an interaction gives rise to an energy gap at the
Fermi level, which makes the usual RKKY model not applicable. The
results show that the effective indirect exchange interaction is
described by a range function which decays exponentially with the
distance between magnetic moments. The interaction is also shown
to depend on whether the two localized moments belong to the same
sublattice or are located in two different sublattices.

\end{abstract}
\pacs{75.70.Ak,75.30.Et,75.30.Hx}

\maketitle

\section{Introduction}

Physical properties of two-dimensional (2D) graphene planes are a
subject of current interest due to their unique transport
properties.\cite{novoselov05,novoselov06,tworzydlo06,morozov06}
Some of these properties originate from peculiarities of the
corresponding electronic structure, or more specifically from the
peculiar Fermi surface of a 2D graphene plane, which consists of
six points located at the corners of the corresponding hexagonal
2D reciprocal lattice.\cite{vincenzo84,saito98} Only two of these
points, however, are nonequivalent. Apart from this, the low
energy excitations at these points can be approximated by the
energy dispersion relation linear in the wavevector. Such electron
states can be described by the Dirac equation (therefore, the
corresponding points of the Brillouin zone are also termed as the
Dirac points). All these features of the electronic structure make
graphene very attractive from both theoretical and experimental
points of view. This is because graphene is an ideal natural
system for testing various theoretical models of 2D electron
transport, including for instance the quantum Hall effect, or the
effects due to the geometrical Berry
phase.\cite{novoselov05,novoselov06}

As concerns magnetic properties of graphene, these are still a
subject of discussion in the relevant literature. Some experiments
indicate on the existence of a spontaneous magnetic moment of the
graphene planes.\cite{han03} However, physical origin of the
spontaneous magnetization is not well understood. Several
distinctly different mechanisms leading to ferromagnetism of
graphene have been proposed, including the instability of the
paramagnetic phase due to electron-electron
interaction,\cite{peres05} indirect RKKY exchange coupling between
magnetic moments of structural defects {\it via} mobile
electrons,\cite{vozmediano05} and others.\cite{nomura06}

The indirect exchange interaction between local magnetic moments
is generally determined by electron excitations near the Fermi
energy. Assuming that the Fermi momentum in a graphene plane is
zero and the excitations are gapless, a non-oscillatory indirect
exchange interaction of the RKKY type has been found in
Ref.~[\onlinecite{vozmediano05}]. Moreover, the corresponding
exchange integral was found to be ferromagnetic, with the range
function decaying as $R^{-3}$ with the distance $R$ between
magnetic moments. However, the intrinsic spin-orbit (SO)
interaction in graphene has been neglected in
Ref.~[\onlinecite{vozmediano05}]. Such an interaction in graphene
opens an energy gap at the Fermi level,\cite{kane05} and makes the
usual RKKY mechanism not relevant due to the absence of electrons
at the Fermi level. Therefore, another mechanism of the indirect
interaction between magnetic moments of structural defects should
be developed.

In this paper we reconsider the indirect exchange interaction in
graphene, taking into account the role of intrinsic SO
interaction. If the SO gap is relatively small, the excitations
across the gap contribute to the indirect exchange
interaction.\cite{bloembergen55,litvinov01} We show that the
exchange coupling is ferromagnetic, and the corresponding range
function decays exponentially with the distance between magnetic
moments -- provided the chemical potential lies in the gap.
Moreover, the spin-spin coupling is shown to be described
approximately by the isotropic three-dimensional (3D) Heisenberg
Hamiltonian. Apart from this, the interaction is shown to be
dependent on the location of magnetic impurities -- the
interaction for impurities located in the same sublattice is
different from that between impurities located in different
sublattices.

\section{Model and exchange interaction}

Indirect exchange interaction between magnetic moments of
structural defects (for instance impurities) is determined by the
electronics structure of the relevant system. To describe electron
states in a graphene plane, we assume the Hamiltonian of
non-interacting electrons in the following form\cite{kane05}:
\begin{equation}
\label{1} H=\hbar v\left( \tau _xp_zk_x+\tau _yk_y\right) +\Delta
\tau _zp_z\sigma _z,
\end{equation}
where three types of the Pauli matrices; $\btau ,\, {\bf p}$, and
$\bsig $, have been introduced. These matrices operate in
different spaces, so their products in Eq.(1) should be understood
as the direct matrix product. The matrix $\btau$ acts in the
(real) space of two nonequivalent sublattices. The matrix ${\bf
p}$, in turn, acts in the (reciprocal) space of two nonequivalent
Dirac points, whereas the  matrix $\bsig$ operates in the electron
spin space. Finally, the constants $v$ and $\Delta$ in Eq.(1) are
the velocity of electrons near the Fermi level (in the limit of no
intrinsic SO interaction), and the parameter of SO interaction,
respectively.

As we have already mentioned above, the Hamiltonian (1) operates
in the 8-dimensional space, and describes electrons in the
vicinity of two nonequivalent Dirac points located at the corners
of the Brillouin zone, ${\bf k}_0$ and $-{\bf k}_0$, with ${\bf
k}_0=\left( 2\pi /3a_0,\, 0\right) $ and $a$ denoting the lattice
constant. The wavevector $k$ in Eq.~1 is measured from the
location of a given Dirac point.

The energy spectrum of the Hamiltonian (1) consists of four-fold
degenerate dispersion curves given by
\begin{equation}
\label{2} \varepsilon _k=\pm \left(\Delta ^2+\hbar^2v^2k^2\right)
^{1/2}.
\end{equation}
As follows from Eq.~(2), the SO interaction opens a gap in the
electronic states around the Fermi level, which extends from
$-\Delta$ to $\Delta$.

We assume that the interaction of 2D electrons with  magnetic
impurities located at ${\bf R}_i$ ($i=1,2,...$) in the graphene
plane can be described by the following Hamiltonian:
\begin{equation}
\label{3} H_i=\frac{g_0}{n}\left[ 1+(-1)^s \tau _z\right] \left( {\bf
S}_i\cdot \bsig \right) \, \delta \left( {\bf r}-{\bf R}_i\right),
\end{equation}
where ${\bf S}_i$ is the spin moment of $i$-th impurity, $g_0$ is
the parameter of exchange coupling between the localized spin and
band electrons, and $n$ is the areal density of the host atoms in
the graphene plane. The parameter $s=0,1$ distinguishes the two (A
and B) sublattices -- if a given impurity belongs to the
sublattice A (B) then $s=0$ ($s=1$). This form of interaction
accounts for the exchange coupling that is sensitive to the
location of the magnetic impurities in the two sublattices.
However, the interaction (3) does not take into account the
intervalley transitions. We consider first this simplified
magnetic interaction, assuming that the rate of intervalley
transitions is small. Later on we will briefly analyze the effect
of intervalley scattering on the indirect exchange interaction.

The Green function corresponding to the Hamiltonian (1) has the form
\begin{equation}
\label{4} G_\varepsilon ({\bf k})=\frac{\varepsilon +\hbar v\left(
\tau _xp_zk_x+\tau _yk_y\right) +\Delta \tau _zp_z\sigma _z}
{[\varepsilon +i\delta \, {\rm sgn}(\varepsilon )]^2-\varepsilon
_k^2}.
\end{equation}
The corresponding Green function in the energy-coordinate
representation can be derived  from the above equation by
integrating $G_\varepsilon ({\bf k})$ over the momentum in the
vicinity of each valley,
\begin{eqnarray}
\label{5} G_\varepsilon ({\pm \bf R})=\int \frac{d^2{\bf k}}{(2\pi
)^2}\; e^{\pm i\left(t{\bf k}_0+{\bf k}\right) \cdot {\bf R}}\,
G_\varepsilon ({\bf k}),
\end{eqnarray}
where $t=\pm\bf 1$ corresponds to the two valleys located at
$\pm\bf k_0$, respectively.

Taking into account Eqs. (4),(5) and following
Ref.~[\onlinecite{dugaev94}] one obtains
\begin{eqnarray}
\label{6} G_\varepsilon ({\pm \bf R}) =e^{\pm it{\bf k}_0\cdot {\bf
R}} \left[ -\frac{i\left( \varepsilon +\Delta \tau _zp_z\sigma
_z\right) }{4\hbar^2v^2} \hskip1cm \right. \nonumber
\\ \left. \times H_0^{(1)}\left( \frac{R\sqrt{\varepsilon
^2-\Delta ^2}}{\hbar v}\right) \pm \frac{\sqrt{\varepsilon
^2-\Delta ^2}}{4\hbar^2 v^2R} \left( R_x\tau _xp_z+R_y\tau
_y\right) \right. \nonumber \\ \left. \times H_1^{(1)}\left(
\frac{R\sqrt{\varepsilon ^2-\Delta ^2}}{\hbar v}\right) \right],
\hskip1cm
\end{eqnarray}
where $H_\nu ^{(1)}(z)$ are the Hankel functions which decay
exponentially in the upper half-plane of the complex argument
$z$.\cite{abramowitz} The Green function defined by Eq.~(5) is the
analytical function in the complex plane of $\varepsilon $, except
for two cuts along the  real axis from $-\infty $ to $-\Delta $ and
from $\Delta $ to $+\infty $.

In a general case, the indirect exchange interaction is mediated
by inter- and intra-valley electron excitations. As already
mentioned above, we consider first the intra-valley contribution
and then discuss the role of inter-valley transitions. In the loop
approximation\cite{dugaev94,litvinov98} we find the interaction
energy of two magnetic impurities ${\bf S}_1$ and ${\bf S}_2$
separated by a distance $R$ in the form
\begin{equation}
\label{7}
E_{int}(R)=w_{\alpha\beta}^{ss^\prime }(R)\,
S_{1\alpha}\, S_{2\beta},
\end{equation}
where
\begin{eqnarray}
\label{8} w_{\alpha\beta}^{ss^\prime}(R)= -\frac{ig_0^2}{n^2}\,
{\rm Tr}\int \frac{d\varepsilon }{2\pi }\, \left[ 1+(-1)^{s}\tau
_z\right] \sigma _\alpha \, G_\varepsilon ({\bf R}) \nonumber \\
\times \left[ 1+(-1)^{s^\prime}\tau _z\right] \sigma _\beta \,
G_\varepsilon (-{\bf R}).
\end{eqnarray}
Making use of Eqs. (5) and (7) and calculating the trace, we find
\begin{eqnarray}
\label{8} w_{\alpha\beta}^{ss^\prime}(R) =\frac{ig_0^2}{\pi
n^2\hbar^4 v^4} \int _{i\delta }^{\infty +i\delta }d\varepsilon
\left\{ \delta_{ss^\prime} \left[ \left( \varepsilon ^2-\Delta
^2\right) \delta _{\alpha \beta} \right. \right. \nonumber \\
\left. \left. +\Delta ^2\delta _{\alpha z}\delta _{\beta z}\right]
\left[ H_0^{(1)}\left( \frac{R\sqrt{\varepsilon ^2 -\Delta ^2}}{v
\hbar}\right) \right] ^2 +\delta_{\alpha\beta}
\left(1-\delta_{ss^\prime}\right)\right. \nonumber \\ \left. \times
\left(\varepsilon ^2-\Delta ^2\right) \left[ H_1^{(1)} \left(
\frac{R\sqrt{\varepsilon ^2-\Delta ^2}}{\hbar v}\right) \right] ^2
\right\} ,\hskip0.5cm
\end{eqnarray}
where the integral is along the real axis above the cut. After
rotating the integration contour to the complex $\varepsilon
$-plane along the imaginary axis, as explained in
Ref.~[\onlinecite{dugaev94}], we find from Eq.~(9)
\begin{eqnarray}
\label{9} w_{\alpha\beta}^{ss^\prime}(R) =-\frac{4g_0^2}{\pi ^3
n^2\hbar^4 v^4 }\int _0^{\infty}d\xi \left\{
\delta_{ss^\prime}\hskip1cm \right. \nonumber \\ \left. \times
K_0^2\left( \frac{R\sqrt{\xi ^2+\Delta ^2}}{\hbar v}\right) \left[
\left( \xi ^2+\Delta ^2\right) \delta _{\alpha \beta} -\Delta
^2\delta _{\alpha z}\delta _{\beta z}\right] \right. \nonumber \\
\left. +\delta _{\alpha\beta}\left(1-\delta_{ss^\prime}\right)
\left( \xi ^2+\Delta ^2\right)
K_1^2\left( \frac{R\sqrt{\xi ^2+\Delta ^2}}{\hbar v}\right)
\right\},
\end{eqnarray}
where $K_\nu (z)$ are the MacDonald functions related to $H_\nu
^{(1)}(z)$ by\cite{abramowitz}
\begin{equation}
\label{10} K_\nu (z)=\frac{i\pi }2\,  e^{i\pi \nu /2} H_\nu
^{(1)}\left( ze^{i\pi /2}\right) .
\end{equation}

When the spin-orbit interaction is taken into account, $\Delta \ne
0$, the range function given by Eq.~(10) scales with the
characteristic length $R_0=\hbar v/\Delta $. The derived exchange
interaction is generally anisotropic. The interaction between
in-plane spin components ($i=x,y$) is given as
\begin{equation}
\label{11} w_{ii}^{ss^\prime }(R) =-\frac{4g_0^2}{\pi ^3 n^2 \hbar v
R_0^3}\left\{\delta_{ss^\prime}I_0(R)
+\left(1-\delta_{ss^\prime}\right)I_1(R)\right\},
\end{equation}
where
\begin{equation}
\label{12} I_{0,1}(R)=\int_0^{\infty}dx \left(
x^2+1\right)K_{0,1}^2\left( \frac{R\sqrt{x^2+1}}{R_0}\right).
\end{equation}
On the other hand, the interaction of the out-of-plane spin
components is expressed {\it via} the formula
\begin{equation}
\label{12} w_{zz}^{ss^\prime }(R) =-\frac{4 g_0^2}{\pi ^3 n^2
\hbar v R_0^3} \left\{
\delta_{ss^\prime}I_2(R)+\left(1-\delta_{ss^\prime}\right)I_1(R)\right\},
\end{equation}
with
\begin{equation}
I_2(R)=\int _0^{\infty}dx\hskip0.1cm x^2 K_0^2 \left(
\frac{R\sqrt{x^2+1}}{R_0}\right).
\end{equation}
The interaction described by Eqs.~(12) and (14) is formally valid
under the condition $R\ge a$, with $a$ being the lattice constant.

The asymptotics of MacDonald functions at large argument,\cite{abramowitz}
 $K_\nu (z)\simeq \sqrt{\pi /2z}\, e^{-z}$, leads to the exponential
 decrease of interactions, $I_{0,1}(R)\sim R^{-3/2}\exp (-2R/R_0)$ and
$I_{2}(R)\sim R^{-5/2}\exp (-2R/R_0)$ for $R\gg R_0$. Thus, the value of
$R_0$ determines an effective range of the exchange interaction.

The magnitude of the indirect exchange interaction depends on the
relative position of the spins in the graphene sublattices. As
mentioned above, the exchange interaction is described by an
anisotropic Heisenberg model. However, $I_0(R)\approx
I_2(R)$\hskip0.1cm, so the anisotropy is weak and one may
approximate the exchange interaction by a 3D-isotropic Heisenberg
Hamiltonian, irrespective which sublattice the magnetic impurities
belong to.

\begin{figure}
\label{fig1}
\includegraphics[angle=-90,width=0.9\columnwidth]{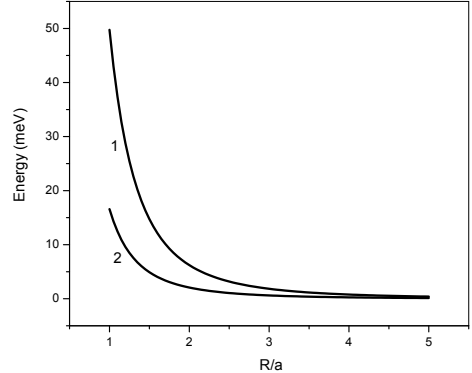}
\caption{Range functions in the presence of spin-orbit
interaction. The curve labelled with 1 represents
$|w_{ii}^{ss\prime}(R)|$, whereas the curve labelled with 2
corresponds to $|w_{ii}^{ss}(R)|$.}
\end{figure}

As follows from Eq.~(10), the gapless fermions ($\Delta =0$)
mediate the indirect exchange interaction that is also
3D-isotropic, but with the power-law range function,
\begin{equation}
\label{13} w_{\alpha\beta}^{ss^\prime }(R) =-\frac{4g_0^2
\delta_{\alpha\beta}}{\pi ^3 n^2 \hbar v R^3} \left\{F_0
\delta_{ss^\prime}+F_1 \left(1-\delta_{ss^\prime}\right)\right\},
\end{equation}
where
\begin{eqnarray}
\label{14} F_0=\int_0^{\infty}x^2 K_0^2(x)dx\approx 0.31,
\nonumber \\
F_1=\int_0^{\infty}x^2 K_1^2(x)dx\approx 0.93.
\end{eqnarray}
The interaction given by Eq.~(16) is the term similar to that
derived in Ref.~[\onlinecite{vozmediano05}]. It should be noted,
however, that the gap due to spin-orbit interaction induces an
exponential decay of the range function, even in samples with low
density of non-magnetic impurities.

The indirect exchange interaction is ferromagnetic. The numerical
results in Fig.~1 show the absolute value of the interaction energy
when the interacting moments are in the same or different
sublattices. In our estimations we used the carbon-carbon distance
$a=1.42$~\AA , the Fermi velocity $\hbar v=5.7$~eV~\AA , $g_0\simeq
1$~eV, \cite{peres05,vozmediano05} and $\Delta \simeq 1.2$~K.
\cite{kane05} From the unit cell area of $S=3a^2\sqrt{3}/2$
containing two carbon atoms, one can estimate the coupling
coefficient $g_0/n\simeq 2.62$~eV~\AA$^2$. In order to illustrate
the range function dependence on $R/a$, we took into account the
scaling coefficient $a/R_0=2.5\times 10^{-5}$.

The numerical results explicitly show the exponential decay of the
range function with the distance between two magnetic impurities.
Moreover, they also demonstrate that the coupling is stronger
when the magnetic moments are localized on different sublattices.

\section{Magnetic polarization due to a single impurity}

The exchange interaction between a magnetic impurity and 2D
electrons gives rise to a spin polarization of the latter ones.
The corresponding spin density distribution ${\bf M}(R)$ can be
calculated as a response of the electron system to the
perturbation due to a single magnetic impurity ${\bf S}_0$ at
${\bf R}=0$. As a result, one finds ${\bf M}(R)$ in the form
\begin{eqnarray}
\label{15} {\bf M}(R)= -ig_0\, {\rm Tr}\int \frac{d\varepsilon
}{2\pi }\, \bsig\, G_\varepsilon ({\bf R}) \left(1+(-1)^{\tau}\tau
_z\right) \nonumber \\ \times \left( \bsig \cdot {\bf S}_0\right)
\, G_\varepsilon (-{\bf R}) .
\end{eqnarray}

The assumed exchange interaction between a magnetic impurity and
2D electrons is local, see Eq.~(3). As a result, the spin
polarization at a given point of the lattice, mediated via the 2D
electrons, is sensitive to the location of the magnetic impurity
(in one sublattice or the other), see Eq.~(16).

Performing calculations similar to those presented above one finds
\begin{equation}
\label{17} M_\alpha (R)=\frac{w_{\alpha \beta }^{s\ne s^\prime
}S_{0\beta }}{g_0}.
\end{equation}
The above result shows that spin polarization has no oscillatory
component. Apart from this, one can conclude from Eqs.~(10) and
(19), that the $z$-component of magnetic density ${\bf M}(R)$ is
nonzero only in the case when the magnetic impurity has a
non-vanishing spin component perpendicular to the graphene plane.

\section{Effect of intervalley transitions}

As we have already mentioned above, graphene has two nonequivalent
valleys in the Brillouin zone, which are associated with two Dirac
points. Moreover, the exchange coupling between localized moment
and 2D band electrons assumed in Eq.~3 does not allow for
intervalley transitions. However, such transitions are allowed by
symmetry of the system and should be taken into account. To do
this we consider an additional term in the interaction between a
localized moment and band electrons. This new term may be written
in the form
\begin{equation}
\label{18} H_{iv}=\frac{g_v}{n}\sum_{i}\left(1+(-1)^s \tau
_z\right) p_x \left( {\bf S}_i\cdot \bsig \right) \, \delta \left(
{\bf r}-{\bf R}_i\right),
\end{equation}
where $g_v$ is the corresponding coupling parameter.

The above form of interaction between 2D band electrons and
localized moments gives rise to additional terms in the indirect
exchange interaction. These terms are of oscillatory type, $\sim
\cos \left( 2{\bf k}_0\cdot {\bf R}\right)$. However, we believe
that the coupling coefficient $g_v$ is much smaller than $g_0$, so
the main results and conclusions described above will not be
changed by including the contribution following from the
interaction (20).

\section{Summary and conclusions}

In this paper we have calculated the indirect exchange interaction
between two magnetic impurities localized in a graphene plane. The
interaction is mediated by two-dimensional electrons and holes,
and acquires the form of an isotropic 3D Heisenberg coupling, with
the coupling parameter being of ferromagnetic type. The exchange
parameter has an exponentially decaying long distance tail
determined by the spin-orbit interaction.

Thus, graphene doped with magnetic impurities may become
ferromagnetic at a certain impurity concentration and below a
certain temperature. Magnetic graphene could be then an ideal
system for experimental investigations of 2D transport of spin
polarized electrons. Moreover, one may also expect that magnetic
graphene could be very useful in spintronics devices, where
electron spin plays a role comparable to its charge.\cite{prinz98}

We have calculated the exchange interaction for $T=0$. Strictly
speaking, our results refer to $T\ll \Delta $. The effect of
temperature on the exchange interaction in narrow-gap semiconductors
has been studied by Rusin,\cite{rusin96} who showed that for the
chemical potential in the gap, the temperature corrections do not
change the main result: even for $T\gg \Delta $, the interaction at
large distances is exponentially decaying.

When calculating the exchange interaction we assumed the chemical
potential $\mu =0$. The results do not change for a finite $\mu $
as long as the chemical potential is located within the energy
gap, which corresponds to the undoped material. In principle, the
chemical potential in graphene can be varied either by a gate
voltage or by doping with impurities. As a result, the chemical
potential can be shifted into the conduction or valence bands, and
the interaction between magnetic impurities includes then the
usual RKKY exchange term {\it via} free carriers. As in
conventional metals or doped semiconductors, this gives an
additional oscillating term with the oscillation period determined
by the Fermi momentum of electrons (holes). However, the mechanism
of exchange interaction considered in this paper is still working
for $|\mu |>\Delta $. The resulting interaction is then due to
both the virtual transitions of electrons through the gap and the
excitations of real electron-hole pairs in the vicinity of the
Fermi energy.

Magnitude of SO interaction in graphene was recently a subject of
extensive discussion in the relevant
literature.\cite{huertas06,min06,yao06} A similar problem was also
discussed in the case of carbon
nanotubes.\cite{ando00,martino02,chico04} Some results indicate
that the SO interaction  in graphene can be very small. However,
the key point is that this interaction is nonzero, which results
in a finite exchange interaction length. As we have shown in this
paper, the nonzero SO interaction leads to some anisotropy of the
magnetic interaction between magnetic ions.

The energy gap in graphene can be also induced by excitonic
effects.\cite{khveshchenko01,gorbar02}, thus being not related to
the spin degrees of freedom. As a result, the exchange interaction
mediated by virtual transitions through the excitonic gap $\Delta
_{exc}$ acquires the finite length $~\hbar v/\Delta _{exc}$. The
magnetic interaction is then strictly isotropic.

\section*{Acknowledgements}
We thank N. M. R. Peres for discussions. This research was
supported by the Portuguese program POCI under Grant No.
POCI/FIS/58746/2004, polish  Ministry of Science and Higher
Education through the Research Project Nr. N202 142 31/2598
(2006-2009), and by the STCU Grant No.~3098 in Ukraine.

\end{document}